\documentclass[11pt,a4paper]{article}
\pdfoutput=1
\usepackage{jcappub}
\usepackage{amsfonts}
\usepackage{wasysym}
\usepackage{amssymb}
\usepackage{epsfig}
\usepackage{textcomp}
\usepackage{graphicx,psfrag}
\usepackage[caption=false]{subfig}
\usepackage{braket}
\usepackage{color}
\usepackage{placeins}
\usepackage{float}
\usepackage{subfig}

\title{A 3.55 keV line from DM $\to a \to \gamma$: predictions for cool-core and non-cool-core clusters}
\author[a]{Joseph P. Conlon,}
\author[a]{Andrew J. Powell}
\affiliation[a]{Rudolf Peierls Centre for Theoretical Physics, University of Oxford,\\
1 Keble Road, Oxford, OX1 3NP, United Kingdom}
\emailAdd{j.conlon1@physics.ox.ac.uk}
\emailAdd{andrew.powell2@physics.ox.ac.uk}

\abstract{We further study a scenario in which a 3.55 keV X-ray line arises from decay of dark matter to an axion-like particle (ALP),
that subsequently converts to a photon in astrophysical magnetic fields. We perform numerical simulations of Gaussian random magnetic fields
with radial scaling of the magnetic field magnitude with the electron density, for both cool-core `Perseus' and non-cool-core `Coma' electron density profiles.
Using these, we quantitatively study the resulting signal strength and morphology for cool-core and non-cool-core clusters. Our study includes the
effects of fields of view that cover only the central part of the cluster, the effects of offset pointings on the radial decline of signal strength and the effects
of dividing clusters into annuli.
We find good agreement with current data and make predictions for future analyses and observations.
}

\arxivnumber{}

\newcommand{\be}{\begin{equation}}
\newcommand{\ee}{\end{equation}}
\newcommand{\bea}{\begin{eqnarray}}
\newcommand{\eea}{\end{eqnarray}}

\newcommand{\ti}{\times}

\newcommand{\mc}{\mathcal}

\newcommand\T{\rule{0pt}{3ex}}       
\newcommand\B{\rule[-2ex]{0pt}{0pt}} 

 \newlength{\wth}
\begin{document}

\maketitle
\flushbottom

\section {Introduction}

Recent results suggest the existence of a 3.55 keV photon line both in a number of galaxy clusters and in the Andromeda galaxy (M31) \cite{Bulbul, Boyarsky}. This line has been observed with both XMM-Newton MOS and PN cameras across a variety of sources at different redshifts, and additionally with both ACIS-I and ACIS-S configurations on the Chandra satellite.
It is absent in blank sky spectra and therefore does not appear to be an instrumental effect such as a slight change in the effective area at $\sim 3.5$ keV. The observations of clusters in \cite{Bulbul} cannot exclude the possibility that the line is an unidentified atomic line, or contamination from an anomalously bright nearby emission line, from the hot ionised intracluster medium (ICM). The simultaneous observation however of the line in M31 in \cite{Boyarsky} argues against this interpretation: a galaxy is a very different environment to a cluster, and unlike clusters, galaxies are not suffused with hot multi-keV gas. While due scepticism is required, the most exciting interpretation of this as yet unidentified line is as arising from the decay or annihilation of dark matter particles.

The canonical decaying dark matter model for this line is a 7.1 keV sterile neutrino decaying as $\tilde{\nu} \to \nu \gamma$ (for example see \cite{Abazajian14}).
Since the observations in \cite{Bulbul, Boyarsky} many other candidates for 7.1 keV dark matter decaying directly to photons have been proposed. While
of considerable theoretical interest, such models are observationally indistinguishable from the sterile neutrino case, as the signal is set purely by the
dark matter density and its decay rate to photons.

The decay rate of a sterile neutrino of mass $m_s$ is set by its mixing angle with the active neutrinos, $\sin^2 \left( 2 \theta \right)$, as
\be
\Gamma_{\tilde{\nu} \to \nu \gamma} = 2.5 \ti 10^{-28} {\rm s}^{-1}  \left( \frac{\sin^2 \left( 2 \theta \right) }{10^{-10}} \right) \left( \frac{m_s}{7.1\ {\rm  keV}} \right)^5.
\ee
Both the observational signals found in \cite{Bulbul, Boyarsky}, and upper bounds found in \cite{Horiuchi, 14057943},
 are expressed in terms of effective inferred values of $\sin^2 \left( 2 \theta \right)$, and for this reason we shall express observational results as effective values of $\sin^2 \left( 2 \theta \right)$ for the discussion of signal strengths.

There are reasons to be dissatisfied with the standard decaying dark matter interpretation. In particular, the dark matter decay rate and hence the inferred value of $\sin^2 \left( 2 \theta \right)$  
differs significantly between observations.
 In \cite{Bulbul}, the signal extracted from a stacked sample of clusters gives $\sin^2 \left( 2 \theta \right) \sim 6 \ti 10^{-11}$,
while observations of Perseus give $\sin^2 \left( 2 \theta \right) \sim 20 - 55 \ti 10^{-11}$ from each of XMM-MOS, Chandra ACIS-I and Chandra ACIS-S observations. 
These two observations are in tension at high statistical significance.
Stacked observations of the local clusters Coma, Ophiuchus and Centaurus with XMM-MOS also give $\sin^2 \left( 2 \theta \right) = 18 \pm 4 \ti 10^{-11}$, and is thus also in mild tension with the stacked cluster observations. 
Observations of M31 prefer a value in the range $\sin^2 \left( 2 \theta \right) =2-20\times 10^{-11}$ \cite{Boyarsky}, where the large range takes into account the uncertainty on the amount of dark matter in M31.
In contrast, the recent null observation of a line at 3.55 keV in the Milky Way corresponds to an upper limit $\sin^2 \left( 2 \theta \right) \lesssim 2.5 \ti 10^{-11}$ at 95\% confidence level \cite{14057943}. This limit rules out the parameter space which explains the observations of \cite{Bulbul} and most of the parameter space of \cite{Boyarsky}, and is thus a significant problem for the standard decaying dark matter interpretation. In addition, the null observation of the Milky Way imposes even more
stringent constraints on models of the line as due to annihilating dark matter, ruling them out by several orders of magnitude.

The analysis from \cite{Boyarsky} includes observations of both M31 and Perseus using offset pointings (centre of the field of view located at some radius from the centre of the cluster). Observations of M31 between $0.4-0.9$ degrees from the centre produce only upper limits on the flux. This decrease in line flux is consistent with the interpretation that the line comes from a dark matter origin. Observations of Perseus grouped into three offset radii also shows a decrease in flux consistent with a dark matter origin, though the uncertainty on the amount of dark matter in Perseus does not allow firm conclusions to be drawn. Stacking the Perseus  observations of \cite{Boyarsky} results in $\sin^2(2\theta) \sim 7-40 \times 10^{-11}$, which is consistent with the values found in \cite{Bulbul} (due to the large uncertainty), but again ruled out by the upper limit of \cite{14057943}. We summarise the observational constraints on  $\sin^2 \left( 2 \theta \right)$ in Table \ref{table:obs}.

\begin{table}[t]
\centering
\begin{tabular}[h]{c | c||c}
\hline
Sample & Instrument& $\begin{matrix}\sin^2(2\theta)\T\B\\(10^{-11})\end{matrix}\T\B$ \T\B\\
\hline
\hline
Full Sample &$\text{XMM}\quad\begin{matrix}\text{MOS}\T\B\\\text{PN}\T\B\end{matrix}$\T\B& $\begin{matrix}6.8^{+1.4}_{-1.4}\T\B\\6.7^{+1.7\ \dagger}_{-1.0}\T\B\end{matrix}$\ \cite{Bulbul}\T\B\\
\hline
$\begin{matrix} \text{Coma + Ophiuchus +}\\\text{Centaurus} \end{matrix}$&$\text{XMM}\quad\begin{matrix}\text{MOS}\T\B\\\text{PN}\T\B\end{matrix}$& $\begin{matrix}18.2^{+4.4}_{-5.9}\T\B\\ <17.6\T\B\end{matrix}$\T\B\ \cite{Bulbul}\\
\hline
Perseus & $\begin{matrix}\text{XMM}\quad\begin{matrix}\text{MOS} \T\B\\\text{PN}\T\B\end{matrix}\T\B\\\text{XMM}\quad\begin{matrix}\text{MOS} \T\B\\\text{PN}\T\B\\\end{matrix}\quad\text{(No Core)}\T\B\\\text{Chandra}\quad\begin{matrix}\text{ACIS-S}\T\B\\ \text{ACIS-I}\T\B\end{matrix}\begin{matrix}\quad\text{(No Core)}\T\B\\ \T\B\end{matrix}\end{matrix}$ & $\begin{matrix}\begin{matrix}55.3^{+25.5}_{-15.9}\T\B\\ <18.8\T\B\\\end{matrix}\T\B\\\begin{matrix}23.3^{+7.6}_{-8.9}\T\B\\ <17.6 \T\B\\\end{matrix}\T\B\\ \begin{matrix}40.1^{+14.5}_{-13.7}\T\B\\ 28.3^{+11.8}_{-12.1}\T\B\end{matrix}\T\B\end{matrix}$\T\B\ \cite{Bulbul}\\
\hline
Perseus (Offset) & XMM-MOS & $\sim 7-40^{\ \dagger\dagger}\ $\cite{Boyarsky}\T\B\\
\hline
`All Other Clusters'& $\text{XMM}\quad\begin{matrix}\text{MOS}\T\B\\\text{PN}\T\B\end{matrix}$& $\begin{matrix}6.0^{+1.1}_{-1.4}\T\B\\5.4^{+0.8}_{-1.3}\T\B\end{matrix}$\ \cite{Bulbul}\T\B\\
\hline
Virgo & $\text{Chandra}\quad\text{ACIS-I}$& $< 10.5$\T\B\ \cite{Bulbul}\\
\hline
M31 (Andromeda) & $ \text{XMM}$& $2.2-20$\T\B\ \cite{Boyarsky}\\
\hline
Milky Way & $\text{Chandra}\quad\text{ACIS-I}$& $\lesssim 2.5$ (95\% C.L.)\ \cite{14057943}\T\B\\
\hline
\end{tabular}
\caption{Observational constraints on $\sin^2(2\theta)$ from \cite{Bulbul,Boyarsky,14057943}. All errors are quoted at 68\% confidence, all upper limits on $\sin^2(2\theta)$ are quoted at 90\% confidence unless stated otherwise. $\dagger$: the line energy in the MOS and PN from the full stacked sample in \cite{Bulbul} are in tension, this $\sin^2(2\theta)$ reduces if the line energies are made to coincide at 3.57 keV. $\dagger\dagger$: the line energy in \cite{Boyarsky} is slightly lower than in \cite{Bulbul}, if the energies were fixed to coincide these $\sin^2(2\theta)$ values would likely go down.}
\label{table:obs}
\end{table}

We focus in this paper on the proposal in \cite{14032370} that the 3.55 keV line signal instead arises from the process ${\rm DM} \to a \to \gamma$: a 7.1 keV dark matter particle decay
produces an axion-like particle (ALP) with $E_a = 3.55 \ {\rm keV}$. This ALP then converts into an $E_{\gamma} = 3.55 \ {\rm keV}$ photon in astrophysical magnetic fields 
through standard ALP-photon oscillations. This scenario is consistent with both
bosonic and fermionic dark matter (including sterile neutrinos). In this model the flux is now also dependent on the probability an ALP converts to a photon before detection, not just the decay rate and density of dark matter. It is thus \emph{expected} that determinations of $\sin^2(2\theta)$ from observations of different objects will be incompatible with each other. The spatial morphology is also quite distinctive from the case of ${\rm DM} \to \gamma$ as
the signal then depends on the magnetic field structure along the line of sight.

The basic qualitative predictions of this scenario were outlined in \cite{14032370}: as $a \to \gamma$ conversion depends on the square of the
transverse magnetic field,
between different clusters the signal will be stronger for clusters with larger magnetic fields, and within clusters the signal will be stronger in the central
region where the magnetic field is larger. As clusters are much larger than galaxies, the $a \to \gamma$ conversion probability is generally much larger in clusters, and
this scenario predicted in \cite{14047741} the subsequent non-observation of the line in the Milky Way \cite{14057943}.\footnote{Current observations imply the M31 regular magnetic field
is both much larger, and coherent over a much larger scale, than the Milky Way \cite{FletcherBeck}, making M31 a much more efficient converter of ALPs to photons and allowing the
 model to be consistent with the observation of the 3.55 keV line in M31 \cite{Boyarsky}.}

This study will purely be concerned with the DM$\rightarrow a \rightarrow \gamma$ signal in galaxy clusters, of which there is the most detailed evidence \cite{Bulbul}. The aim of this paper is to make the above statements about galaxy clusters more quantitative, by analysing the propagation of a
3.55 keV ALP through numerically simulated magnetic fields that roughly correspond to the types of magnetic field structure expected in actual
galaxy clusters.
In particular we focus on the differences between DM$\rightarrow a \rightarrow \gamma$ and DM$\rightarrow \gamma$, and within  DM$\rightarrow a \rightarrow \gamma$ we distinguish between observations of `cool-core' and `non-cool-core' clusters (see Section \ref{sec:clusters}).

The basic ALP Lagrangian is
\begin{equation}
\mathcal{L} =  \frac{1}{2} \partial_{\mu} a \partial^{\mu} a - \frac{1}{2} m_a^2 a^2 + \frac{a}{M} {\bf E} \cdot {\bf B}.
\label{eqn:alpL}
\end{equation}
Throughout this paper we assume a massless ALP with $m_{a} = 0$. The general properties and phenomenology of ALPs are reviewed in \cite{Ringwald},
and the physics of $a \to \gamma$ conversion in the magnetic fields of galaxy clusters is described in detail in
\cite{13053603, 13123947, 14032370}.

In the next section we review clusters and outline the difference between cool-core and non-cool-core clusters. We discuss the computational methodology in Section \ref{sec:methodology}. We then present our quantitative results in Section \ref{sec:results}, and finally we conclude in Section \ref{sec:conclusions} by stating our predictions for the model, in light of future planned analyses.

\section{Galaxy Clusters}
\label{sec:clusters}

This section mostly contains a standard review on galaxy clusters relevant for the ${\rm DM} \to a \to \gamma$ scenario, particularly the distinction between cool-core and non-cool-core clusters. It is included
for completeness as it may be unfamiliar to readers with a particle theory background.

Galaxy clusters are the largest virialised structures in the universe involving $100-1000$ galaxies and a characteristic size of $\mc{O}(1) $ Mpc.
The dominant mass contribution is dark matter, which makes up $ \sim 90\%$ of the mass of the cluster. The baryonic matter consists principally of a hot ionised
plasma --- the intracluster medium (ICM) --- with characteristic temperature $T \sim 2 - 10 \ {\rm keV}$ and free electron density $n_e \sim 10^{-3}\ 
{\rm cm}^{-3}$. This hot gas gives rise to large-scale diffuse X-ray emission via thermal bremsstrahlung. The ICM is also permeated by turbulent multi-scale magnetic fields,
with $\mc{O}(1 - 10)$ kpc coherence lengths. The magnetic fields are strongest in the center of a cluster and
decrease towards the outskirts with a scaling $B(r) \sim B_0 \left( \frac{n_e(r)}{n_e(0)} \right)^{\eta}$.
Faraday rotation measurements imply that the typical size of the cluster magnetic field is $\mc{O}(\mu {\rm G})$, with values up
to $\mc{O} (10\ \mu {\rm G})$ at the centre of cool-core clusters \cite{Feretti}.

The diffuse X-ray emission implies that the ICM is losing energy and will over time, if isolated, cool.
As the cooling is collisional, the cooling will occur most rapidly in the centres of clusters where the ICM is densest.
The apparent cooling time in the centre of dense clusters is much less than the age of the cluster, and
it was originally believed --- the `cooling flow' model --- that for an isolated cluster the cooling was runaway,
 giving rise to cold gases and intense star formation at the centre of the cluster. However this is not observed, and it
is now understood that this does not occur and a feedback mechanism, possibly associated to central AGN activity,
 injects energy into the cooling core leading to a steady state.

Clusters can be divided into two distinct types: cool-core clusters and non-cool-core clusters (for example, see \cite{09110409}). Roughly, a cool-core cluster can be viewed as the endpoint of the above evolution. Cool-core clusters are characterised by a central core region within which the ICM temperature decreases
sharply, the ICM electron density increases sharply,
and the ICM magnetic field increases sharply. The size of the core region is rather small compared to the entire cluster (e.g. $R_{core} \lesssim 100 \ {\rm kpc}$
compared to $R_{cluster} \sim 1 \ {\rm Mpc}$). A prototypical example of a cool-core cluster is Perseus, for which deep Ms Chandra observations of the core are described in
\cite{astroph0510476}.

If a cluster has been disturbed --- for example through a merger with another cluster --- the hot gas will be disturbed and heated up in the collision (a famous example is the
Bullet cluster). If observed at this epoch in its evolution, the cluster will not show a cool-core, and instead may be roughly isothermal and with a large-scale radio halo.
Such clusters do not have a core region of lower ICM temperatures, and instead show a roughly constant electron density and temperature over the central $\sim 100 $ kpc.
A prototypical non-cool-core cluster is the Coma cluster, which is believed to have undergone a major merger in the relatively recent past.

The core regions of cool-core clusters involve significantly enhanced magnetic fields (observational estimates give $B \sim 10 - 50 \ \mu {\rm G}$, in contrast to $B \sim 1 - 5\ 
\mu {\rm G}$ for the centres of non-cool-core clusters).
 In the context of the ${\rm DM} \to a \to \gamma$ scenario considered here, this implies
the possibility of a significantly enhanced line signal in the central region of cool-core clusters (over and above that coming simply from the higher dark matter density).

\section{Numerical methodology}
\label{sec:methodology}
In this section we briefly describe the procedure used to caculate the flux expected from DM $\to a \to \gamma$. The detailed numerical methodology for generating the magnetic field and calculating probabilities for an ALP to convert to photon in a such a field is described in detail in \cite{13123947} and we refer the reader there for a full description.

Briefly, the magnetic field is taken to be a Gaussian random field with a Kolmogorov power spectrum for scales between $\Lambda_{min}$ and
$\Lambda_{max}$.
The field is generated over a $(1\text{ Mpc})^2 \times 2\text{ Mpc}$ box at 1 kpc resolution. In order to simulate fully along the line of sight (i.e. the z direction), this is taken as the side of length 2 Mpc. The $\begin{pmatrix}x, & y\end{pmatrix}$ plane is then a $1 \text{ Mpc}^2$ square defined perpendicular to the line of sight. Since the simulated magnetic field is isotropic, and to allow maximal radial range in our analysis, the centre of the cluster is taken as a corner of this square. This allows us to go out to radial distances in excess of 1 Mpc.

The generated random field is then modulated such that the magnetic field strength scales with the electron density $n_e$ as $B(r) \sim B_0 \left( \frac{n_e(r)}{n_0} \right)^{\eta}$.
We consider two values for
$\eta$: $\eta = 0.5$ and $\eta = 1$. The former is the best-fit value found for the scaling of the magnetic field with radius in the Coma cluster \cite{Bonafede} (with a best-fit central
magnetic field $B_0 = 4.7\ \mu {\rm G}$).
The latter was found to describe the scaling of the magnetic field with radius in the cool-core cluster Hydra A \cite{KucharEnsslin} (with a best-fit central magnetic field
$B_0 = 36\ \mu {\rm G}$). Thus real fields should fall somewhere in between these two extreme cases.
Note, the larger the value of $\eta$, the sharper the central peak in the magnetic field.

Our main aim is to analyse the difference specifically attributable to the presence of a cool-core spike in the electron density. 
Thus we use the same pre-scaled magnetic field to generate two different magnetic fields corresponding roughly to cool-core (Perseus-like) and non-cool-core (Coma-like) clusters, where the only difference then is the electron density distribution and the central magnetic field value. To generate this field we take $\Lambda_{min} = 4 \ {\rm kpc}$ and $\Lambda_{max} = 10 \ {\rm kpc}$. (Any correlations between magnetic field coherence length and the presence or absence of a cool-core
would be interesting to study, but are beyond the scope of this paper.)

These spectra are \emph{not} the actual magnetic fields of the Coma or Perseus clusters.
For any actual cluster, the range of scales in the magnetic field will differ and the spectrum may not be Kolmogorov (to mention just two of the many
ways this represents an idealisation of the actual cluster magnetic fields). 
However, the above does have a similar form to spectra that capture the Faraday Rotation Measure
observations of various clusters, and it produces multi-scale magnetic fields with approximately the right size of coherence length.
We therefore regard the above as sensible toy models for the magnetic field in cool-core and non-cool-core clusters.

We shall consider two electron densities. For the Perseus-like model, we take the measured electron density for Perseus \cite{ChurazovForman}
\begin{equation}
n_e(r) = \frac{3.90 \times
10^{-2}\ \text{cm}^{-3}}{\left(1+(r/80\ \text{kpc})^2 \right)^{1.8}} +
\frac{4.05 \times 10^{-3}\ \text{cm}^{-3}}{\left(1+(r/280\ \text{kpc})^2
\right)^{0.87}}\, ,\label{nebeta}
\end{equation}
a so-called double-$\beta$ model, and for the Coma-like cluster we take
\begin{equation}
n_e(r) = \frac{3.44 \ti 10^{-3}\text{ cm}^{-3}}{\left( 1 +  (r/291 \text{ kpc})^2\right)^{1.125}}
\end{equation}
based on the $\beta$-model for Coma \cite{Briel}.

The probability that the produced ALP converts to a photon before it leaves the cluster is computed numerically using the full equation of motion for the coupled ALP-photon system. 
Whilst it does not give the full result, it is illuminating to consider the analytic solution for a single coherent magnetic field domain, as it will inform us of the expected results of the full simulation.
For the ALP energies we are considering the probability per domain is
\be
P(a\rightarrow \gamma)\propto \left(\frac{B_{\perp}(r)}{M}\right)^2,
\ee
with $M$ defined in equation \ref{eqn:alpL}. Since $B\propto n_e(r)^{\eta}$, it is a decreasing function of $r$, and the relative morphology between cool-core and non-cool-core clusters is set by the electron density. A steep decline of $n_e$ with radius will translate into a very steep radial morphology for the conversion probabilities. This will be the origin of the difference between cool-core and non-cool-core clusters.

Since we want to isolate the effects of cool-core vs non-cool-core, we use the same dark matter distribution for both clusters. We take an NFW profile \cite{NFW}
\be
\rho(r)=\rho_0\frac{1}{(r/r_s)(1+r^2/r^2_s)}
\ee
with radial scale $r_s = 360\ {\rm kpc}$ and concentration parameter $c_{200}=5$. We discretize this profile capping the density such that it is flat over the central 1 kpc, to avoid the density blowing up at the $r = 0$ grid point. The flux is then calculated by summing the product $\rho(x,y,z)\times P_{a\rightarrow \gamma}(x, y, z)$ at each grid point, where $P_{a\rightarrow\gamma}(x,y,z)$ is the probability that an ALP, created from dark matter decay at position $\begin{pmatrix}x,& y,&z\end{pmatrix}$, travelling along the line of sight, converts to a photon before leaving the simulated volume of the cluster.

\section{Results}
\label{sec:results}
In the process ${\rm DM} \to a \to \gamma$, the magnitude of the signal depends on the product $\Gamma_{DM \to a} \ti P(a \to \gamma)$. As the former is unknown, and the latter depends
on $\frac{1}{M^2}$, which is also unknown, it is not possible to make useful predictions for the absolute magnitude of the signal (although it is possible to make predictions for the relative magnitude between each cluster, since $ P(a \to \gamma)\propto B^2$).
We shall therefore not focus on the overall normalisation of the signal, but instead on the morphology and in particular on how the inferred value of
$\sin^2 \left( 2 \theta \right)$ would vary between observations.

One of the most noticeable features of the data presented in \cite{Bulbul} is the much stronger signal arising in the centre of the Perseus cluster --- and also to a
slightly lesser extent in the stacked sample of local clusters: Coma, Ophiuchus and Centaurus --- compared to the stacked sample of distant clusters, see Table \ref{table:obs}.
 For local clusters the line signal is extracted from the full field of view (FOV) of XMM-Newton, since these clusters fill the FOV and it is not possible to consider the full cluster in a single central pointing.
For more distant clusters, the entire cluster can be contained
within the FOV and the line signal is extracted from within the virial radius of the cluster $r_{500}$, with a typical
extraction radius of 1 Mpc.

While this makes no difference for the case of ${\rm DM} \to \gamma$, where dark matter produces photon lines at a constant rate independent of
its location in the cluster, this has a significant effect on ${\rm DM} \to a \to \gamma$. As the magnetic field strength
decreases towards the cluster outskirts, with a scaling $B \sim n_e(r)^{\eta}$, it is clear that a line signal extracted within the central
250 kpc would be automatically stronger than one extracted from within the central Mpc. It then follows that a line signal extracted from nearby clusters that
do not fit fully within the FOV will be systematically higher than a line signal extracted from distant clusters where the entire cluster can be put in the FOV.
We will use our simulations to quantify the expected differences between near and far clusters in Section \ref{sec:nearvfar}.

The other unique feature of the DM$\rightarrow a \rightarrow \gamma$ model is the radial morphology. There are two ways of analysing signal morphology. First, there is the absolute morphology: the surface brightness of the 3.55 keV line signal as a function of radial distance from the
centre of the cluster. This will peak in the centre of the cluster, where both the dark matter density and magnetic field are greatest. However, this is also true for
the conventional scenario of ${\rm DM} \to \gamma$. We therefore focus more on the relative morphology: this is the relative strength of the
 ${\rm DM} \to a \to \gamma$ signal compared to a ${\rm DM} \to \gamma$
signal. 
The relative morphology is then propotional to the value of $\sin^2(2\theta)$ that would be inferred from that observation. As such we would expect a constant relative morphology across all observations in the case where dark matter decays directly to photons.
The fact that for DM$\rightarrow a \rightarrow \gamma$ the relative signal strength, or inferred value of  $\sin^2(2\theta)$ increases towards the centre is purely a feature of the magnetic field morphology of the cluster, which increases in magnitude towards the centre. Again we quantify this feature in the following sections, \ref{sec:offsetpoint} and \ref{sec:radialbehaviour}.

\subsection{Nearby vs Distant Clusters}
\label{sec:nearvfar}
We first give results for the comparative effect of observing nearby or distant clusters. The data in \cite{Bulbul} involves three separate sub-samples:
Perseus, Coma + Ophiuchus + Centaurus, and `all other clusters'. Perseus is at a redshift of 0.018 and the XMM-Newton FOV at this redshift
corresponds to a radius of 250 kpc (15 arcminute radius). In \cite{Bulbul} Perseus was also observed with Chandra ACIS-I and ACIS-S configurations.
The former involves the four chips I0, I1, I2, I3 with the extraction from a circular FOV (8 arcminutes = 136 kpc radius) and the latter the single chip S3
(square FOV with side length 8 arcminute), without the central 1 arcminute. The extraction radii for Centaurus, Coma and Ophiuchus with XMM-Newton were 170 kpc, 330 kpc and
410 kpc. The stacked `all other clusters' sample is at an averaged redshift of $z=0.06$, with 42 clusters having an extraction radius $> 1 \ {\rm Mpc}$.

\begin{figure}[t]
\centering
	\subfloat[]{
		\includegraphics[width=11cm]{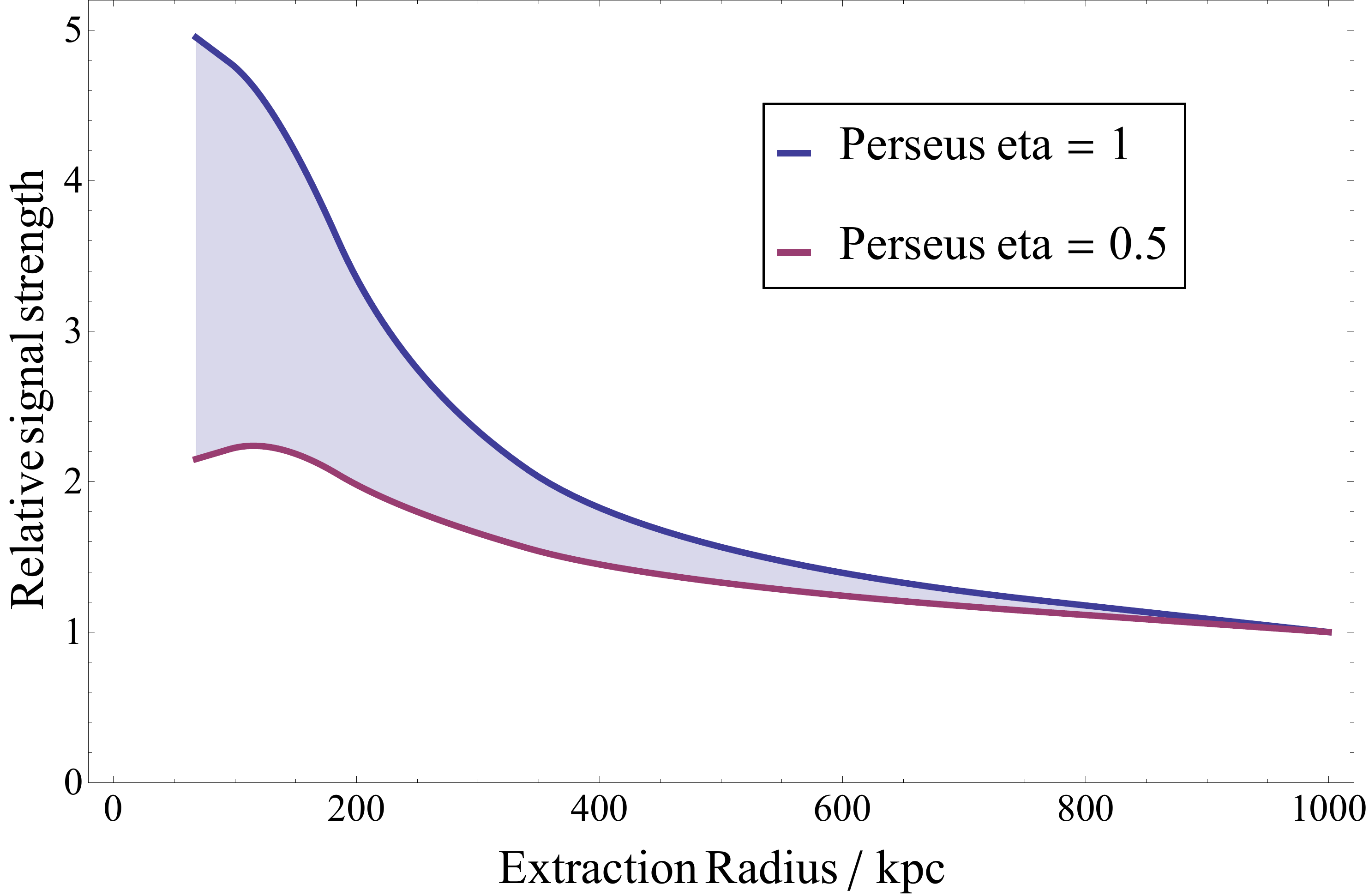}
	}
	\\
	\subfloat[]{
		\includegraphics[width=11cm]{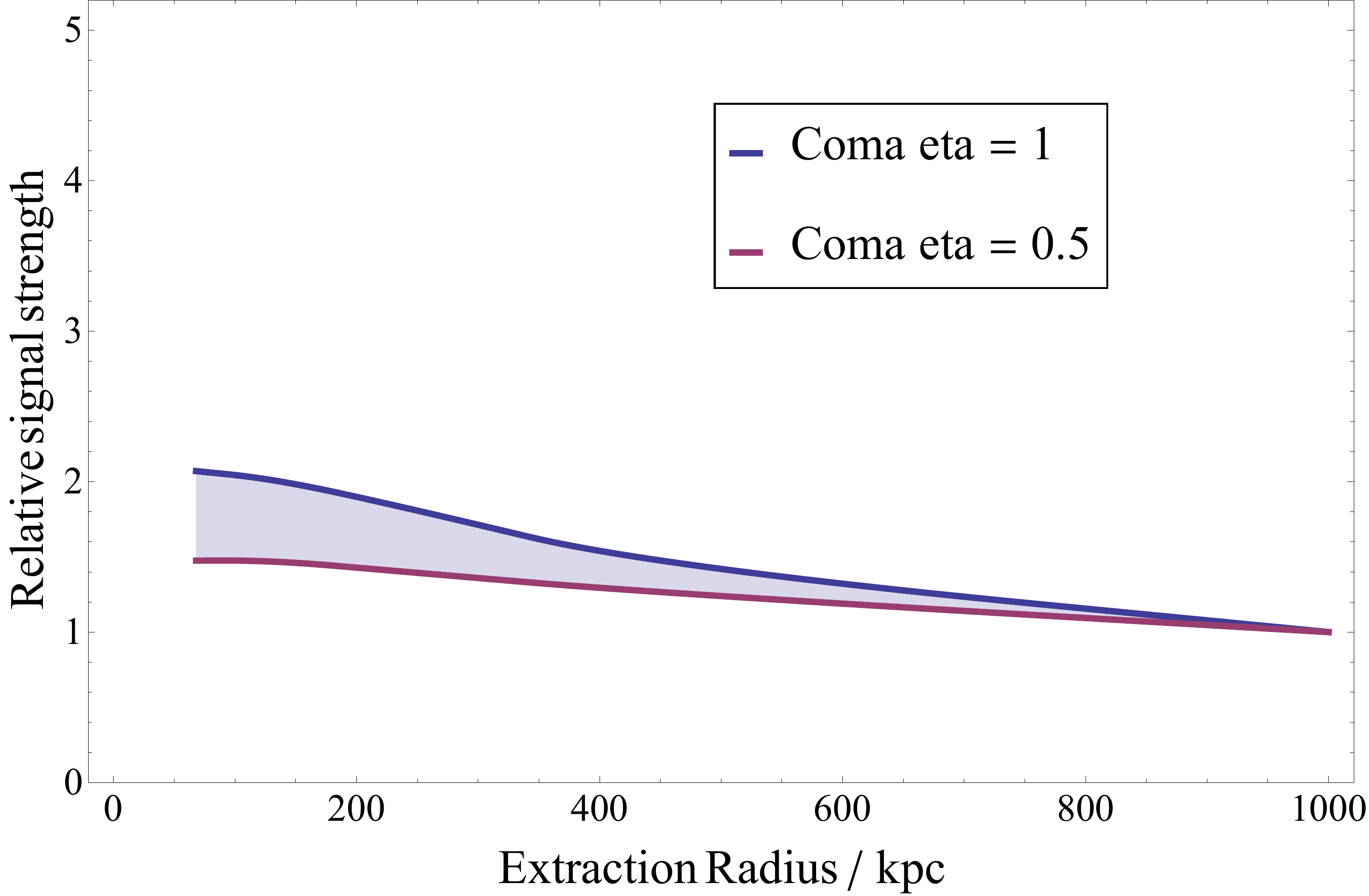}
	}
	\caption{The relative signal strength for the a) cool-core cluster and b) non-cool-core cluster as a function of extraction radius.
The relative strength --- the inferred value of $\sin^2 \left( 2 \theta \right)$ --- is plotted relative to that for an extraction radius of $1\ {\rm Mpc}$, which has been set to have strength of unity. We note that DM$\rightarrow \gamma$ would have a normalised relative signal strength of 1 for all field of views.}
\label{fig:extractionradii}
\end{figure}

In Figure \ref{fig:extractionradii} we plot the relative signal strength (proportional to the effective $\sin^2(2\theta)$) obtained for various extraction radii.
All are taken to be perfectly on-center observations.
The strength is normalised so that the signal for an extraction radius of 1 Mpc is one (in arbitrary units).
 We plot this for each of the four models described above: `Perseus' with $\eta = 0.5$ and
$\eta = 1$, `Coma' with $\eta = 0.5$ and $\eta = 1$. This shows the enhancement in the
inferred value of $\sin^2 \left( 2 \theta \right)$ from observing local clusters where only the central
region can fit in the telescope FOV.

We observe an increase in inferred value of $\sin^2 \left( 2 \theta \right)$ as the extraction radius shrinks.
We see that the effect is particularly marked for cool-core clusters --- the `Perseus' case --- where there is a notable
 change in signal strength with extraction radius.
For such clusters, extraction of the signal within a central $\sim 200\ {\rm kpc}$ radius leads to a value of $\sin^2 \left( 2 \theta \right)$
that can be $2-4.5$ times stronger
than for a signal extracted over an entire $1\ {\rm Mpc}$ radius.
For non-cool-core clusters, the effect is less pronounced: for the `Coma' case with $\eta = 0.5$ there is only a
forty per cent difference between $R_{ext} = 100 \ {\rm kpc}$ and $R_{ext} = 1$ Mpc.

This is interesting in light of the results of \cite{Bulbul}. The Coma + Ophiuchus + Centaurus sample is at an averaged redshift of 0.02, with Perseus at $z=0.016$, compared to the stacked sample, which is at an averaged redshift of 0.06. The average extraction radius for the Coma + Ophiuchus + Centaurus sample is then between $200-400$ kpc, whereas the `all other clusters' sample has an average extraction radius of $\sim 1000$ kpc. Our prediction of an increase of about a factor of $1.5-4.5$ appears to be borne out in the data of \cite{Bulbul} (see Table \ref{table:obs}). The Coma + Ophiuchus + Centaurus sample has a $\sin^2(2\theta)$ that is $\sim 3$ times that derived from the `all other clusters' sample. Similarly the value derived from Perseus is $\sim 4$ times that of the `all other clusters' sample. 

It is also interesting to compare our predictions to the data from Perseus. The line signal in Perseus has been observed in two different satellites, and with three different fields of view, and is thus the perfect place to test our prediction. The XMM-MOS observations have the largest FOV radius of 250 kpc, neglecting the central 20 kpc, the signal gives $\sin^2(2\theta)\approx 23.3$. The observation with the smallest FOV is with the Chandra ACIS-S instrument which also excises the central 20 kpc, the signal corresponds to $\sin^2(2\theta)\approx 40.1$. This is a factor of $\sim 1.7$ increase which is consistent with the increase simulated for a Perseus-like cool-core cluster with a steep radial magnetic field decrease, see Figure \ref{fig:extractionradii}. Similarly, it is expected that the inferred value for $\sin^2(2\theta)$ from the Chandra ACIS-I configuration (FOV radius 136 kpc) should fall inbetween these two values, indeed this is borne out in the data: $\sin^2(2\theta)\approx 28.3$. While there is considerable noise in the data, we thus conclude that the current status of observations of Perseus supports this prediction of our model.

\subsection{Offset Pointing}
\label{sec:offsetpoint}
For the Perseus cluster, the combination of results in \cite{Bulbul} and \cite{Boyarsky} show that the line in Perseus has been
positively observed both with central pointings of
XMM-Newton and also with offset pointings up to an offset radius of $\sim 800 \ {\rm kpc}$.
There is also a data point in \cite{Boyarsky} at $\sim 1.6 \ {\rm Mpc}$ --- however, as this has a flux that
is consistent at 1 sigma with zero, we do not count it as a detection.
 It is therefore interesting to study the expected behaviour of the line signal with offset radius.

\begin{figure}[t]
\centering
	\subfloat[]{
\includegraphics[width=11cm]{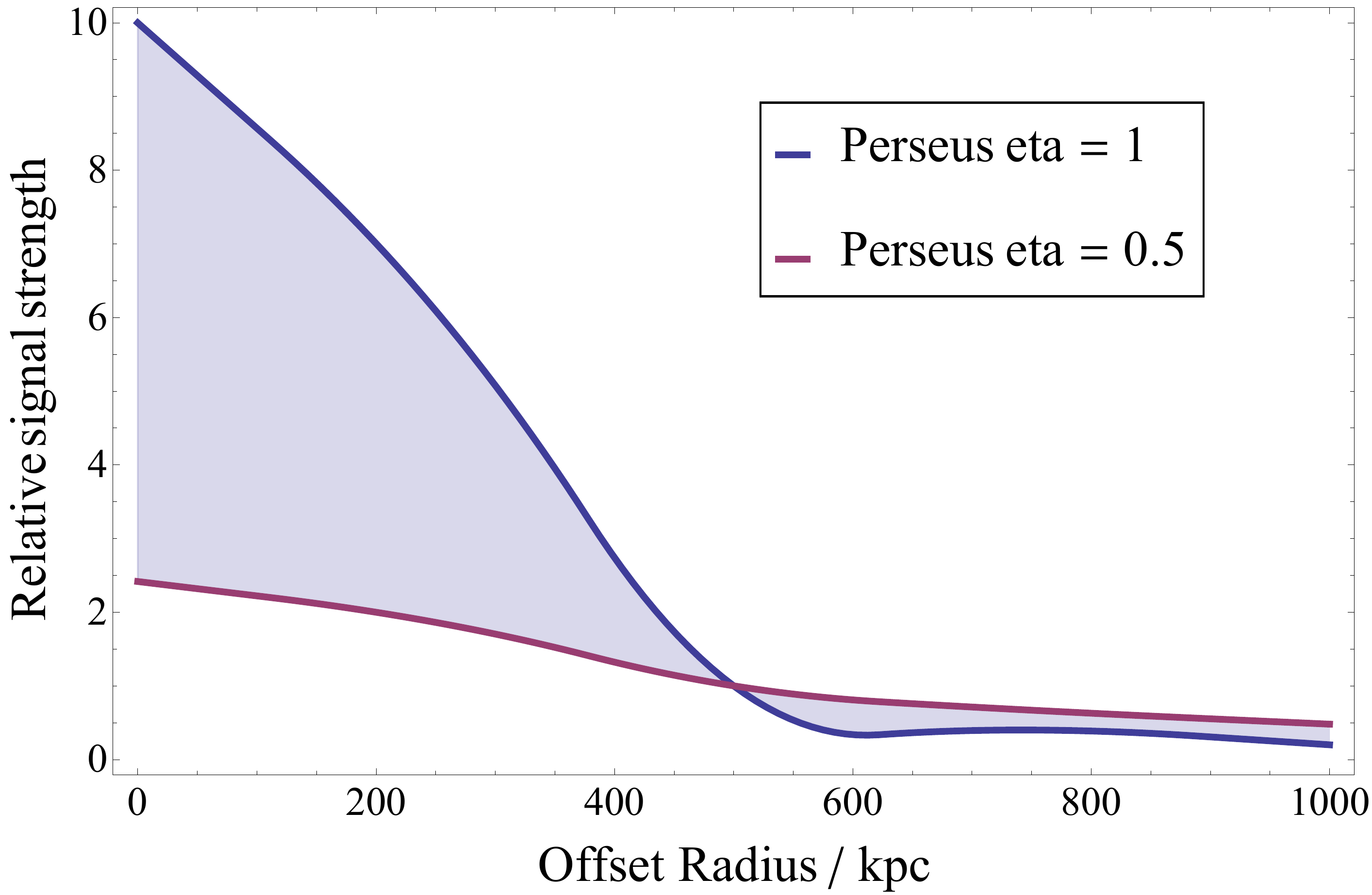}
	}
	\\
	\subfloat[]{
\includegraphics[width=11cm]{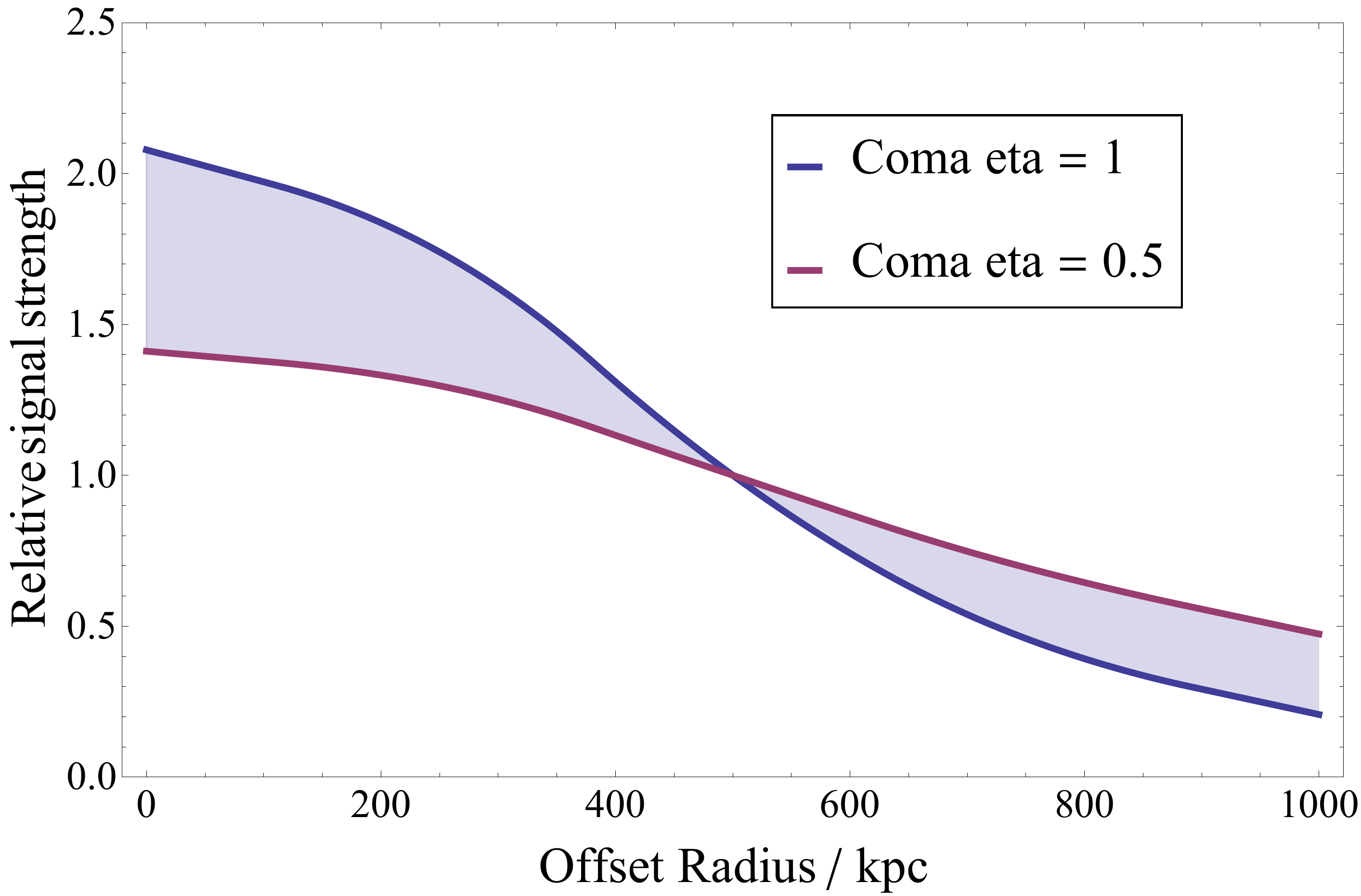}
}
	\caption{The relative strength of the signal for a 250 kpc extraction radius as a function of the offset pointing for a) the cool-core cluster and b) the non-cool-core cluster. The strengths have been normalised so that
an offset pointing 0.5 Mpc from the cluster centre has strength unity. We note that DM$\rightarrow \gamma$ would have a normalised relative signal strength of 1 for all offset radii.}
\label{fig:offset}
\end{figure}

Motivated by the above observations of Perseus we now consider offset pointings, with extraction of the signal
from within a complete FOV of 250 kpc radius.
In Figure \ref{fig:offset}, we plot the relative value of the inferred $\sin^2 \left( 2 \theta \right)$ that would be observed, in the case that the signal is extracted from a
250 kpc radius centred a distance $0 < R < 1000\ {\rm kpc}$ away from the cluster center.

The signal at an offset radius of $500\ \text{kpc}$ is normalised to unity.
Compared to a 500 kpc offset, we see that in general the central pointing leads to a signal that is a factor of a
few stronger. For the case of a cool-core cluster with $\eta = 1$, the enhancement in $\sin^2 \left( 2 \theta \right)$ can be as large as a factor of 10.
Note that the central enhancement is much larger than for the case of Figure \ref{fig:extractionradii}. This is because in Figure \ref{fig:extractionradii} the central core region, with its maximal flux,
is included for all choices of extraction radii. In Figure \ref{fig:offset} however the offset pointings for $r \gtrsim 250 \ {\rm kpc}$ involve no contribution from the central region.

With the data available it is not possible to test this prediction of the model. The large uncertainties in the amount of dark matter in Perseus results in significant  errors on the inferred $\sin^2(2\theta)$ values. We hope that future off-axis observations will shed light on this interesting aspect of the model.

\subsection{Radial Behaviour}
\label{sec:radialbehaviour}
\begin{figure}[t]
\centering
\subfloat[]{
\includegraphics[width=11cm]{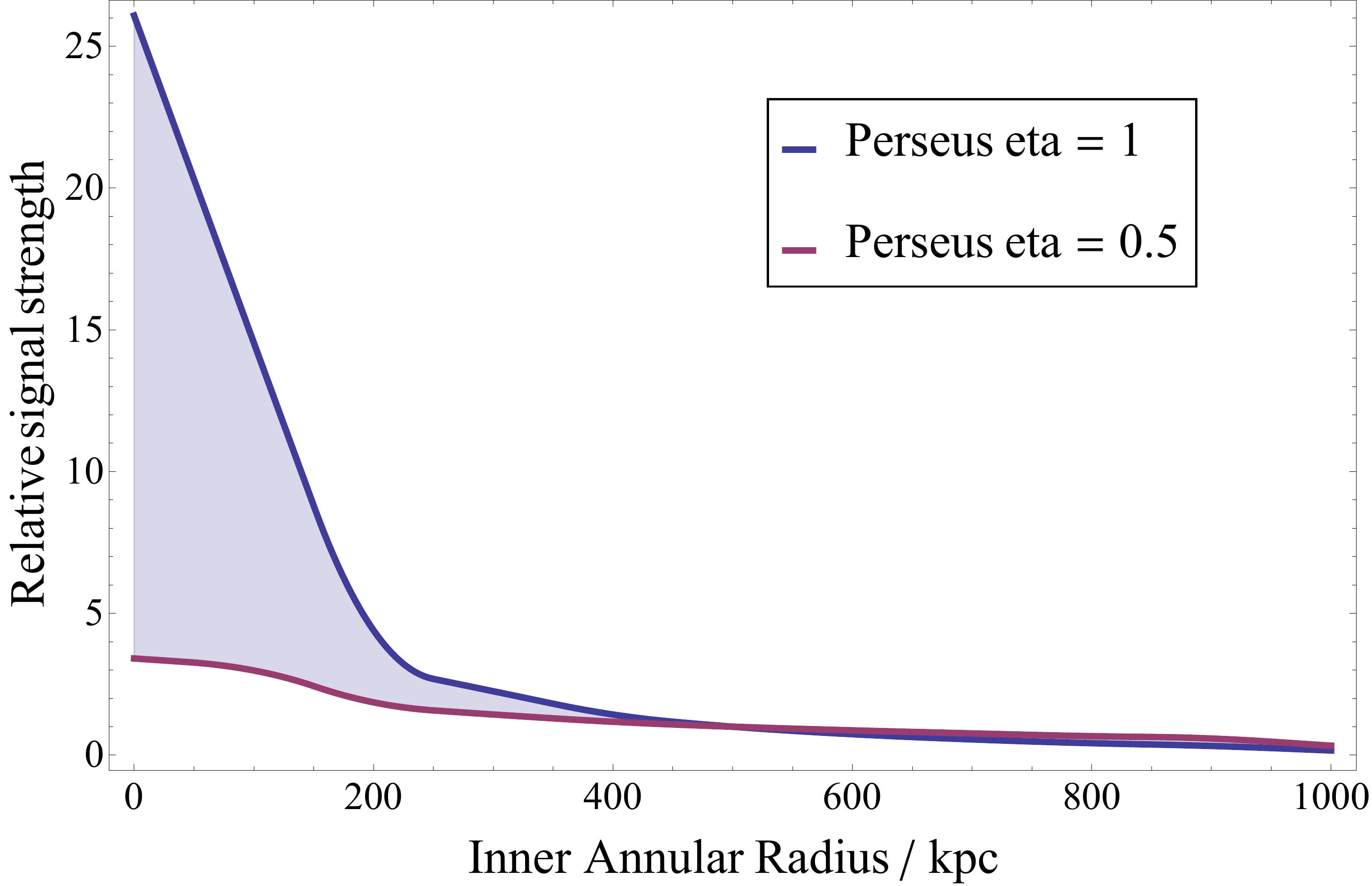}
}
\\
\subfloat[]{
\includegraphics[width=11cm]{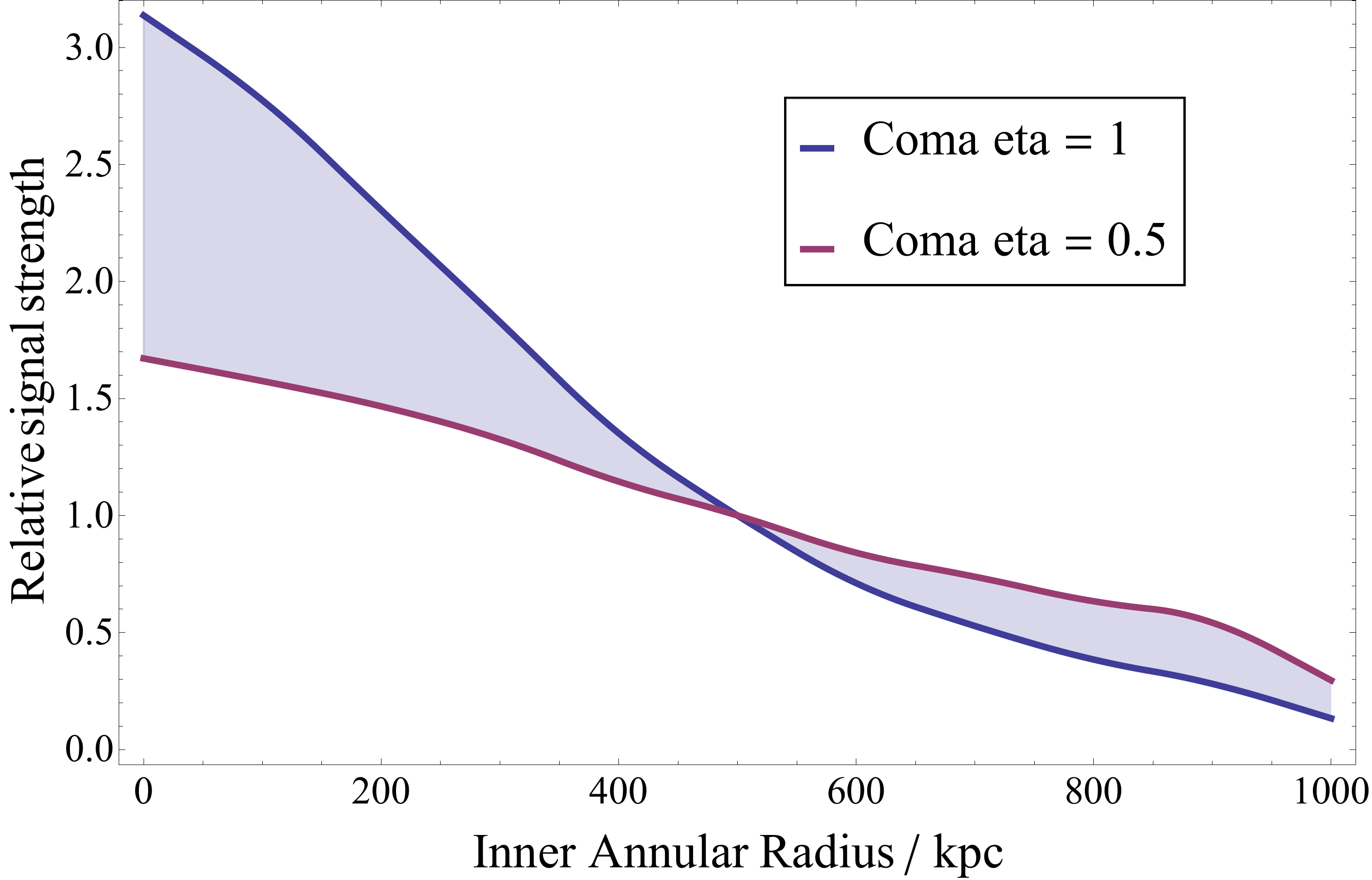}
}
	\caption{The relative strength for different annuli, between $n$ and $(n+1)$*100 kpc, for a) the cool-core cluster, and b) the non-cool-core cluster. The signal strength is normalised so the 500 - 600 kpc
annulus has strength unity. We note that DM$\rightarrow \gamma$ would have a normalised relative signal strength of 1 for all annuli.}
\label{fig:annuli}
\end{figure}

We next consider the effect of dividing clusters into annuli, and extracting the signal separately within each annulus.
According to \cite{Bulbul} this is currently being performed on the large sample of clusters contained there, so one can anticipate
additional data on this in the relatively near future. For concreteness we separate the annular radii in 100 kpc steps.
The results for this are shown in Figure \ref{fig:annuli}.

This captures similar physics to the effect of considering offset pointings, and so the figures have a similar form.
One clear aspect of this is the much sharper central peaking of the signal for the `Perseus' case of cool-core clusters compared to the
`Coma' case of non-cool-core clusters.

We finally make the obvious point that as cool-core clusters have much higher central magnetic fields than non-cool-core clusters, the signal
in their central region will be proportionately larger (we do not know of any reason for cool-core clusters to have larger magnetic fields than non-cool-core
clusters outside the core region). Assuming that the magnetic field at a radius of 500 kpc is roughly comparable, we expect that signals extracted within the
central $\sim 250 \ {\rm kpc}$ will be stronger for cool-core clusters by a factor $1.5 - 4.5$. Thus it would be interesting to stack cluster centres separately as cool-core and non-cool-core in order to see whether this effect is present.

\section{Caveats and Conclusions}
\label{sec:conclusions}
Let us discuss some possible caveats to the above.

One potential systematic error in the above results is that on theoretical grounds there is expected to be a tendency for the coherence lengths to increase towards
the outskirts of clusters, as the electron density decreases and the characteristic physical scale of the problem increases.
 As the conversion probability per domain, of size $L$, grows as $L^2$ (and so the total probability grows as $L$), this will cause a higher signal at large radii, causing the peaking behaviour at low radii in Figures \ref{fig:offset} and \ref{fig:annuli} to be slightly less extreme.

Another potential systematic error lies in our approximation for the magnetic field, which is clearly an idealisation.
One intriguing feature of the results of \cite{Bulbul} is that the effective value of $\sin^2 (2 \theta)$ doubles for the XMM-Newton central pointing including the central
arcminute, compared to the case where the central arcminute is removed. As described in \cite{Bulbul} the conservative interpretation of this is the
additional cold gas in the centre affects the ability to measure the continuum background, leading to a spurious increase in the 3.55 keV line flux.
However there remains the possibility that this is a genuine physical effect, and the 3.55 keV line really has a very sharp peak in the centre of the Perseus
cluster. Although the $\eta =1 $ cool-core model does give a sharp central peaking, even this is insufficient to allow half the flux of the 3.55 keV line to originate
within a $\sim 20 \, {\rm kpc}$ radius of the centre. The actual centre of the Perseus cluster is known from deep Chandra observations to be a very complex and active location \cite{astroph0510476} and any such structure there is by construction not present in our simulations, which assume a smooth scaling of the magnetic field with the electron density.
If this structure leads to an even sharper peak in the magnetic field, then in principle this could give rise to such a sharp central signal.

Based on the above analyses, 
let us list the predictions for the ${\rm DM} \to a \to \gamma$ scenario:
\begin{enumerate}
\item
For a cool-core cluster, the effective value of $\sin^2 \left( 2 \theta \right)$ extracted from the central $ \sim 200 $ kpc
 will be larger by a factor $2-4.5$ than the signal extracted from the central 1 Mpc. For a non-cool-core cluster, the similar enhancement would be a factor $1.4 - 2$.

\item
For a cool-core cluster, the effective value of $\sin^2 \left( 2 \theta \right)$ inferred from an offset pointing at a radial
distance of 500 kpc and with a FOV of 250 kpc is between 2.5 and 10 times smaller than for a central pointing with
a 250 kpc FOV. For a non-cool-core cluster, the difference is in the range $1.4-2$.

\item
Assuming that the `average' magnetic field at a radius of 500 kpc is roughly the same for cool-core and non-cool-core clusters,
and assuming a 250 kpc extraction radius, central pointings on a cool-core cluster should give values of $\sin^2 \left( 2 \theta \right)$
between 1.5 and 5 times larger than central pointings for non-cool-core clusters.

\item
The inferred value of $\sin^2 \left( 2 \theta \right)$ extracted from a stacked sample of local clusters that fill the telescope
FOV will be higher than the inferred value of $\sin^2 \left( 2 \theta \right)$ from a stacked sample of distant clusters, for which the
entire cluster can be put in the telescope FOV.
\end{enumerate}

Our results have most weight when applied to stacked samples of cool-core and non-cool-core clusters, as for any individual cluster there may be uncertainty in a factor
of a few for the absolute value of the magnetic field, depending on the size of the cluster and its history.
These differences will be averaged away by stacking and allow a focus on the typical properties of cool-core and non-cool-core clusters.

It is hoped future observations of the 3.55 keV line will further illuminate its properties, and show whether these predictions are correct.

\section*{Acknowledgments}

JC is funded by a Royal Society University Research Fellowship and by the ERC Starting Grant `Supersymmetry Breaking in String Theory'.
AP is funded by an STFC studentship. We thank Pedro Alvarez, Stephen Angus, Peter Biermann, Esra Bulbul, Francesca Day, David Marsh, Lance Miller, Markus Rummel and Takayuki Tamura for discussions, and the organisers and participants of the Chalonge Meudon Workshop 2014.

\bibliography{refs}
\bibliographystyle{JHEP}

\end{document}